\documentclass[epj]{svjour}
\usepackage{graphics}
\newcommand{\be}{\begin{equation}}
\newcommand{\ee}{\end  {equation}}
\newcommand{\ba}{\begin{eqnarray}}
\newcommand{\ea}{\end  {eqnarray}}
\newcommand{\eq}[1]{equation (\ref{eq:#1})}
\newcommand{\bm}{\bf m}
\newcommand{\bs}{\bf s}
\begin{document}
\title{Structure of the Hilbert-space of the infinite-dimensional
       Hubbard model}
%\subtitle{}
\author{Claudius Gros \and Wolfgang Wenzel
%\thanks{\emph{Present address:} Insert the address here if needed}
}                     % Do not remove
\institute{Institut f\"ur Physik, Universit\"at Dortmund, 
           44221 Dortmund, Germany}
\date{Received: date / Revised version: date}
\abstract{
An iterative procedure for the explicit construction of the nontrivial
subspace of all symmetry-adapted configurations with non-zero weight
in the ground-state of the $\infty$-dimensional Hubbard model is
developed on the basis of a symmetrized representation of the
transition operators on a sequence of Bethe-Lattices of finite
depth. The relation ship between these operators and the well known
mapping of the $\infty$-dimensional Hubbard model onto an effective
impurity problem coupled to a (self-consistent) bath on
non-interacting electrons is given. As an application we
calculate the properties of various Hubbard stars and give
estimates for the half-filled Hubbard model with up to $0.1\%$
accuracy.
} %end of abstract
\maketitle
%%%%%%%%%%%%%%%%%%%%%%%%%%%%%%%%%%%%%%%%%%%%%%%%%%%%%%%%%%%%
\section{Introduction}
\label{intro}
The Hamiltonian of the half-filled Hubbard model
is given by
\begin{eqnarray*}
H\ =\
{-t^*\over\sqrt{2Z}}\sum_{\langle i,j\rangle,\sigma}
f_{i,\sigma}^\dagger f_{j,\sigma}^{\phantom{\dagger}}
\qquad\qquad\qquad\qquad\qquad\\
\qquad+\,U\sum_i
(f_{i,\uparrow}^\dagger f_{i,\uparrow}^{\phantom{\dagger}}
-1/2)
(f_{i,\downarrow}^\dagger f_{i,\downarrow}^{\phantom{\dagger}}
-1/2),
\end{eqnarray*}
where $Z$ is the coordination number of the lattice, i.e.\
$Z=2D$ on a simply hypercubic lattice. The model is
well defined and nontrivial in the limit $D\to\infty$
\cite{Metzner89}. In this limit all correlations 
are local \cite{Metzner89,MH89} and the Hubbard model can
be solved iteratively by mapping it to an
effective Anderson impurity model \cite{Georges92},
which reads for the $n^{\mbox{\small th}}$ iteration as
\begin{eqnarray}
H^{(n)}\ =\ 
\sum_{l=1,\sigma}^{M_\sigma^{(n-1)}} 
\left[ V_{l,\sigma}^{(n-1)}
f_{\sigma}^\dagger c_{l,\sigma}^{\phantom{\dagger}}
+ \mbox{H.C.}\right] \qquad\qquad
\label{H_n}\\
+\sum_{l=1,\sigma}^{M_\sigma^{(n-1)}} \epsilon_{l,\sigma}^{(n-1)}
c_{l,\sigma}^\dagger c_{l,\sigma}^{\phantom{\dagger}}
+U\,
(f_{\uparrow}^\dagger f_{\uparrow}^{\phantom{\dagger}}
-1/2)
(f_{\downarrow}^\dagger f_{\downarrow}^{\phantom{\dagger}}
-1/2),\quad
\nonumber
\end{eqnarray}
where $n=0,1,2\dots$ denotes the number of iterations.
The $f_\sigma^\dagger$ is the creation operator of the
central (impurity) site and $M_\sigma^{(n-1)}$ the number
of state of the bath obtained from the previous ($n-1$)
iterations.
The `onsite' energies $\epsilon_{l,\sigma}^{(n-1)}$
and hybridization matrix elements $V_{l,\sigma}^{(n-1)}$
may be obtained from the local (impurity) Green's function
of the previous iteration, $G_\sigma^{(n-1)}(\omega)$. For the
$\infty$-dimensional Bethe lattice this relation is given
by \cite{Gros94}
\begin{equation}
G_\sigma^{(n-1)}(\omega)\ =\
\sum_{l=1}^{M_\sigma^{(n-1)}} 
{2(V_{l,\sigma}^{(n-1)}/t^*)^2
\over
\omega-\epsilon_{l,\sigma}^{(n-1)}}.
\label{self}
\end{equation}
The spin-dependence, $\sigma=\uparrow,\downarrow$,
has been explicitly retained in above formulas,
in order include the antiferromagnetically ordered
state. 

Once the parameters $\epsilon_{l,\sigma}^{(n-1)}$ and
$V_{l,\sigma}^{(n-1)}$ have been extracted from
(\ref{self}) one needs to calculate
$G_\sigma^{(n)}(\omega)$ by solving (\ref{H_n}).
The relevant Hilbert-space is of the order of
\begin{equation}
M_\sigma^{(n)} \approx
4\cdot 2^{M_\uparrow^{(n-1)}}\cdot 2^{M_\downarrow^{(n-1)}}.
\label{M_n}
\end{equation}
The number of poles in $G_\sigma(n)(\omega)$
will of the same order of magnitude as (\ref{M_n}), 
exponentially larger than $M_\sigma^{(n-1)}$.
In any numerical treatment of it is therefore necessary
to throw away an exponentially large part of the Hilbert-space
\cite{Georges96,Bulla98}.

Above route to solve the infinite-D Hubbard model is
aimed to determine the one-particle Green's function
and it is not obvious from (\ref{H_n}) how to
construct the eigenstates in the original Hilbert-space,
in particular of the ground-state wavefunction.
Here we show, that it is possible to construct
the ground-state wavefunction {\it iteratively},
taking the half-filled Bethe lattice as an
example. 

%%%%%%%%%%%%%%%%%%%%%%%%%%%%%%%%%%%%%%%%%%%%%%%%%%%%%%%%%%%
\begin{figure}
\resizebox{0.50\textwidth}{!}{
\includegraphics{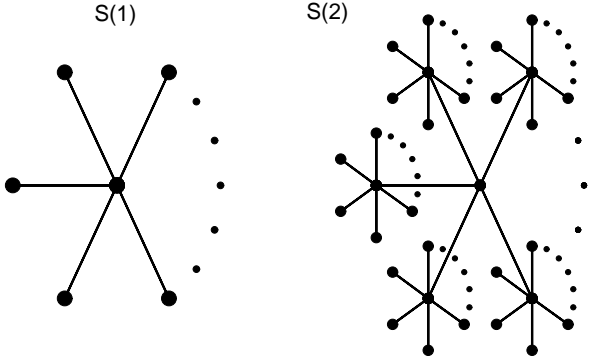}
}
\caption{Illustration of the Hubbard star S(1) and the
         star of the stars S(2).}
\label{Fig1}     
\end{figure}
%%%%%%%%%%%%%%%%%%%%%%%%%%%%%%%%%%%%%%%%%%%%%%%%%%%%%%%%%%%

%%%%%%%%%%%%%%%%%%%%%%%%%%%%%%%%%%%%%%%%%%%%%%%%%%%%%%%%%%%%
\section{Transition Operators on the Hubbard-Star}
\label{Hilbert-space}

A natural approach to solve
(\ref{H_n}) iteratively is to consider
the series of generalized Hubbard stars \cite{Dongen91},
which are truncated Bethe-lattices.
We denote with S(0) the single site, with
S(1) the Hubbard star, with S(2) the
star of stars, etc., see Fig.\ \ref{Fig1}.
This approach has previously been
applied successfully to the study of the
Mott-Hubbard transition \cite{Gros94}
and allows to determine the ground-state
of the infinite-D Hubbard model iteratively.
In following we restrict ourselves to the case
of the antiferromagnetic state at half filling,
generalization to other fillings and frustrated
models are straightforward.

We write the Hamiltonian of S(n) as
\begin{eqnarray}
H^{(n)} \ =\ \sum_{i=1}^Z H_i^{(n-1)} -
{t^*\over\sqrt{2}}\sum_{\sigma=\uparrow,\downarrow}
\left[
f_{\sigma}^\dagger A_{\sigma}^{\phantom{\dagger}}
+A_{\sigma}^\dagger f_{\sigma}^{\phantom{\dagger}}
\right]\nonumber\\
+U\,
(f_{\uparrow}^\dagger f_{\uparrow}^{\phantom{\dagger}}
-1/2)
(f_{\downarrow}^\dagger f_{\downarrow}^{\phantom{\dagger}}
-1/2),\qquad\quad
\label{S_n}
\end{eqnarray}
where the $f_\sigma^\dagger$ are the electron creation
operators at the central site and
\begin{equation}
A_\sigma^\dagger\ =\ {1\over\sqrt{Z}}\sum_{i=1}^Z c_{i,\sigma}^\dagger.
\label{A}
\end{equation}
Here the $c_{i,\sigma}^\dagger$ are the electron creation operators
of the respective central sites of S$_i$(n-1). The $H_i^{(n-1)}$ 
in (\ref{S_n}) are
the Hamiltonians of the S$_i$(n-1), $i=1,\dots Z$.

We will show next that the knowledge of the
exact eigenstates of $H_i^{(n-1)}$ allows for
the construction of the exact eigenstates
of $H^{(n)}$ with a finite number of operations,
despite the fact that $Z\rightarrow\infty$
stars of order $(n-1)$ couple to the central site.

We now introduce a notation for the eigenstates of
$H_i^{(n-1)}$.
Let $|k_{\bm},i\rangle$ designate the k-th state in the sector 
${\bm} = (m_+,m_-)$ containing $m_\sigma$-particles of spin-$\sigma$ on
subsystem i. We measure $m_\sigma$ relative to the 
ground-state of $H_i^{(n-1)}$.
For the enumeration of the accessible states in the Hilbert space it
is useful to define many-body transition operators
\be
 |k_{\bm + \bs}^{\phantom{\prime}}\rangle
 \langle k_{\bm}^\prime|_i = 
\bigotimes_{j < i} {\bf 1}_j \,  \otimes\,
|k_{\bm + \bs}^{\phantom{\prime}},i\rangle\langle i,k_{\bm}^\prime| \,
\otimes \, \bigotimes_{i < j} {\bf 1}_j  
\ee
for the subsystem $i$ with $\bs = (1,0)$ or $\bs = (0,1)$ 
respectively. We note that these operators obey the 
anti-commutation relations
\ba
 && \big\{\,
    |k_{\bm + \bs^\prime}^{\phantom{\prime}}\rangle
    \langle k_{\bm}^\prime|_i \, ,\,
    |q_{\bm + \bs}^{\phantom{\prime}}\rangle
    \langle q_{\bm}^\prime|_j
\,\big\}\  = \qquad\quad\nonumber \\
 && \hspace*{1cm} \delta_{i,j} 
   \,  \delta_{k_{\bm}^\prime,q_{\bm + \bs}^{\phantom{\prime}}}  
   |k_{\bm + \bs^\prime}^{\phantom{\prime}}\rangle
   \langle q_{\bm}^\prime|_i \label{icommute}\\
 && \hspace*{0.6cm} +\ \delta_{i,j} 
   \,  \delta_{q_{\bm}^\prime,
     k_{\bm + \bs^\prime}}  
   |q_{\bm + \bs}^{\phantom{\prime}}\rangle
   \langle k_{\bm}^\prime|_i 
\nonumber
\ea
Since $A_\sigma^\dagger$ creates only symmetrized combinations of states on
the periphery of the star, it is useful to rewrite $A_\sigma^\dagger$ as
\be
A_\sigma^\dagger = \sum 
a(k_{\bm + \bs}^{\phantom{\prime}},k_{\bm}^\prime)\,
|k_{\bm + \bs}^{\phantom{\prime}}\rangle
\langle k_{\bm}^\prime| 
\label{eq:adagger}
\ee
where we introduced the symmetrized transition operator
\be
 |k_{\bm + \bs}^{\phantom{\prime}}\rangle
\langle k_{\bm}^\prime| =
  \frac{1}{\sqrt{Z}} \sum_i 
|k_{\bm+\bs}^{\phantom{\prime}}\rangle
\langle k_{\bm}^\prime|_i
\ee
The coupling matrix elements in \eq{adagger} are given as 
\be
 a(k_{\bm+\bs}^{\phantom{\prime}},k_{\bm}^\prime) 
\equiv
\langle k_{\bm + \bs}^{\phantom{\prime}},i | 
c_{i,\sigma}^\dagger |i,k_{\bm}^\prime\rangle,
\ee
independent of $i$.

Using Eq.\ (\ref{icommute}), commutation rules for the
symmetrized operators are easily derived:
\ba
 \big(\, |k_{\bm+\bs}^{\phantom{\prime}}\rangle
\langle k_{\bm}^\prime|\, \big)^2 & = & 
\frac{1}{Z} \sum_{i,j} 
|k_{\bm+\bs}^{\phantom{\prime}}\rangle
\langle k_{\bm}^\prime|_i
|k_{\bm+\bs}^{\phantom{\prime}}\rangle
\langle k_{\bm}^\prime|_j \nonumber \\
  & = & 0. \label{zero}
\ea
Similarly we find:
\ba
 |k_{\bm}^\prime\rangle\langle k_{\bm + \bs^\prime}| 
  | q_{\bm+\bs}^{\phantom{\prime}}\rangle
\langle q_{\bm}^\prime| & = &
 \frac{1}{Z}\sum_{ij} 
  |k_{\bm}^\prime\rangle
\langle k_{\bm + \bs^\prime}|_i
  | q_{\bm+\bs}^{\phantom{\prime}}\rangle
\langle q_{\bm}^\prime|_j  \nonumber\\
 & = & \frac{1}{Z} 
\delta_{k_{\bm + \bs^\prime},q_{\bm+\bs}^{\phantom{\prime}}}  
   \sum_i  |k_{\bm}^\prime\rangle
  \langle q_{\bm}^\prime|_i \nonumber  \\
 & = &\frac{1}{\sqrt{Z}}\  
\delta_{k_{\bm + \bs^\prime},q_{\bm+\bs}^{\phantom{\prime}}}  
|k_{\bm}^\prime\rangle
\langle q_{\bm}^\prime|. \label{anti}
\ea
Note that operators of type 
$
|k_{\bm}^\prime\rangle \langle q_{\bm}^\prime|
$, which do not change the number of particles
on periphery, are bosonic in nature, while the
$
|k_{\bm+\bs}^{\phantom{\prime}}\rangle \langle k_{\bm}^\prime|
$
are fermionic in nature, see (\ref{icommute}).

%%%%%%%%%%%%%%%%%%%%%%%%%%%%%%%%%%%%%%%%%%%%%%%%%%%%%%%%%%%%
\begin{figure}
\resizebox{0.50\textwidth}{!}{
\includegraphics{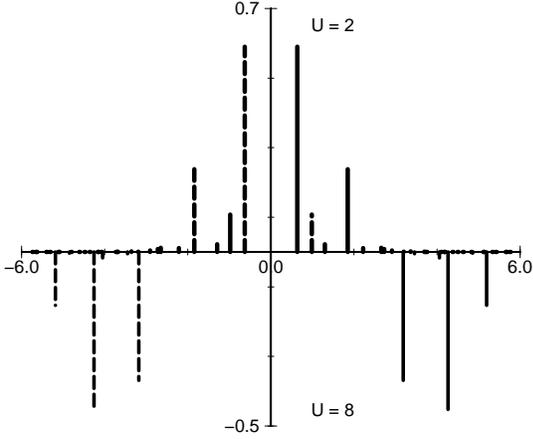}
}
\caption{Distribution of the 
$|a(k_{\pm\bs},0_{(0,0)})|^2 $ for S(2) 
as a function of $t^*$. Solid bars: $\bs=(0,\pm1)$,
dashed bars: $\bs=(\pm1,0)$. For $U=2t^*$ $+|a|^2$
are shown, for $U=8$ $-|a|^2$ are shown. 
The formation of the lower and upper Hubbard
band for $U=8t^*$ can be seen. Here $t^*=1$
has been used. 
        }
\label{Fig2} 
\end{figure}
%%%%%%%%%%%%%%%%%%%%%%%%%%%%%%%%%%%%%%%%%%%%%%%%%%%%%%%%%%%

%%%%%%%%%%%%%%%%%%%%%%%%%%%%%%%%%%%%%%%%%%%%%%%%%%%%%%%%%%%%
\section{Construction of the Hilbert Space}
\label{ground-states}

We define the vacuum $|0\rangle$ of S(n) by
\begin{equation}
 |0\rangle = |0\rangle_{\rm c} \bigotimes_i |0_{(0,0)},i\rangle,
\label{vacuum_0}
\end{equation}
where $|0\rangle_{\rm c}$ 
designates the vacuum of the central site and
$|0_{(0,0)},i\rangle$ the (many-body)
ground-state on the i-th leg of the 
peripheral sites. We define
\begin{equation}
 |\sigma\rangle =
f_\sigma^\dagger\, |0\rangle,\qquad\quad
 |\uparrow\downarrow\rangle =
f_\uparrow^\dagger f_\downarrow^\dagger \,
|0\rangle \label{vacuum_updn}
\end{equation}

In order to iteratively construct the overall Hilbert 
space with nonzero coupling to the vacuum state 
defined above, we concentrate on the accessible states. 
Suppose for simplicity that the central site is occupied 
by an electron with spin
$\sigma=\,\uparrow$. 
The Hamiltonian $H^{(n)}$ then
couples the state $|\uparrow\rangle$ 
directly to the normalized states
\begin{equation}
   |k_{(1,0)}\rangle\langle 0_{(0,0)}|\, |0\rangle
\quad\mbox{and}\quad
|k_{(0,-1)}\rangle\langle 0_{(0,0)}|\, |\uparrow\downarrow\rangle.
\label{first}
\end{equation}
The matrix elements of above states with $|\uparrow\rangle$ 
are 
\[
-t^* a(k_{(1,0)},0_{(0,0)})/\sqrt{2} 
\quad\mbox{and}\quad
-t^* a(k_{(0,-1)},0_{(0,0)})/\sqrt{2}
\]
respectively, i.e.\ they are of order one in the
limit $Z\to\infty$.

To construct the next set of states in the hierarchy of accessible
configuration, we must act with $A_\sigma^{\phantom{\dagger}}$ on the
periphery and re populate the central site.  Without loss of generality
we consider $\sigma=\,\downarrow$.  For the first state in Eq.\
(\ref{first}), $ |k_{(1,0)}\rangle\langle 0_{(0,0)}|\, |0\rangle$, we
have only one possibility,
\begin{equation}
   |q_{(0,-1)}\rangle\langle 0_{(0,0)}| \,
   |k_{(1,0)}\rangle\langle 0_{(0,0)}|\, |\downarrow\rangle,
\label{second_1}
\end{equation}
which is a normalized state. The
normalization of
\begin{eqnarray}
|q_{(1,-1)}\rangle\langle k_{(1,0)}| \,
|k_{(1,0)}\rangle\langle 0_{(0,0)}|\, |\downarrow\rangle
= \qquad\quad
\label{second_2}
\\
\qquad\quad
{1\over\sqrt{Z}}
|q_{(1,-1)}\rangle\langle 0_{(0,0)}|\,|\downarrow\rangle
\nonumber
\end{eqnarray}
is $1/Z$ and therefore vanishes in the
limit $Z\to\infty$ (here we have used
(\ref{anti})). States with overturned spins
on the {\it same} subsystem therefore do
{\it not} couple to the ground-state of
$H^{(n)}$. Generalizing this result one
can show that only states on the periphery with
\begin{equation}
|\,m_\uparrow|\,+\,|\,m_\downarrow\,|\ \le\ 1
\label{delta_m}
\end{equation}
on the same subsystem contribute 
to the ground-state of $H^{(n)}$.

The other state,
$ |k_{(0,-1)}\rangle\langle 0_{(0,0)}|\, 
|\uparrow\downarrow\rangle$,
in Eq.\ (\ref{first}) couples
to $|0\rangle$ and to the normalized state
\begin{equation}
 |q_{(0,1)}\rangle\langle 0_{(0,0)}|\, 
 |k_{(0,-1)}\rangle\langle 0_{(0,0)}|\, |\uparrow\rangle.
\label{second_3}
\end{equation}
The normalization of the state
\begin{eqnarray}
|q_{(0,0)}\rangle\langle k_{(0,-1)}| \,
|k_{(0,-1)}\rangle\langle 0_{(0,0)}|\, |\uparrow\rangle
= \qquad\quad
\label{second_4}
\\
\qquad\quad
{1\over\sqrt{Z}}
|q_{(0,0)}\rangle\langle 0_{(0,0)}|\,|\downarrow\rangle
\nonumber
\end{eqnarray}
does again vanish for $Z\to\infty$. 

We therefore conclude
that of all states on the periphery with
$m_\uparrow=0=m_\downarrow$ only the
ground-state $|0_{(0,0)}\rangle$
couples to the ground-state of $H^{(n)}$.
The complete Hilbert space is therefore
spanned by 
\begin{equation}
\prod_{\{k_{(\pm1,0)}\}}
|k_{(\pm1,0)}\rangle\langle 0_{(0,0)}|
\prod_{\{k_{(0,\pm1)}\}}
|k_{(0,\pm1)}\rangle\langle 0_{(0,0)}|\, 
|x\rangle,
\label{all_states}
\end{equation}
where $|x\rangle=|0\rangle$, 
$|\sigma\rangle$ or
$|\uparrow\downarrow\rangle$ and where
all $\,\{k_{(\pm1,0)}\}\,$ and
$\,\{k_{(0,\pm1)}\}\,$ are mutually distinct, due
to Eq.\ (\ref{zero}). 

%%%%%%%%%%%%%%%%%%%%%%%%%%%%%%%%%%%%%%%%%%%%%%%%%%%%%%%%%%%%
\section{Mapping to the Anderson Model}
\label{Hamiltonian}

Let us define
with $M_{\bs}^{(n-1)}$ the number
of non-zero matrix elements
$a(k_{\bs},k_{(0,0)})$.
The $M_{\bs}^{(n-1)}$ are related
to the $M_\sigma^{(n-1)}$ occurring 
in (\ref{H_n}) via
\begin{equation}
M_\uparrow^{(n-1)} =
M_{(1,0)}^{(n-1)}+ M_{(-1,0)}^{(n-1)}
\label{M_up}
\end{equation}
and
\begin{equation}
M_\downarrow^{(n-1)} =
M_{(0,1)}^{(n-1)}+ M_{(0,-1)}^{(n-1)}.
\label{M_down}
\end{equation}
The hybridization matrix elements
$V_{l,\sigma}^{(n-1)}$ of (\ref{H_n})
are given by
\begin{equation}
V_{l,\uparrow}^{(n-1)}\ =\ {-t^*\over\sqrt{2}}\,
a(l_{(\pm1,0)},0_{(0,0)}),
\label{V_sigma}
\end{equation}
and respectively for $\sigma=\,\downarrow$.
The onsite energies
$\epsilon_{l,\sigma=\uparrow}^{(n-1)}$ of (\ref{H_n})
are given by
\begin{equation}
\pm\epsilon_{l,\uparrow}^{(n-1)}\ =\ 
E(l_{(\pm1,0)})- E(0_{(0,0)}),
\label{epsilon_sigma}
\end{equation}
where the 
$E(l_{\bs})$ are the diagonal energies
of $H_i^{(n-1)}$,
\begin{equation}
E(l_{\bs}) \ \equiv\
\langle l_{\bs},i| H_i^{(n-1)} |i,l_{\bs}\rangle,
\label{E_s}
\end{equation}
independent of $i$. Note, that all energies are
measured with respect to the Fermi-energy $U/2$, which
is absorbed in Eq.\ (\ref{H_n}) in the interaction term.

Let's us see how things work out for S(1) and S(2). 
The half-filled ground state of S(0) is single
occupied, let's say with an up-electron. We have then
\[
M_{(1,0)}^{(0)} = 0 = M_{(0,-1)}^{(0)},\qquad
M_{(-1,0)}^{(0)} = 1 = M_{(0,1)}^{(0)}
\]
with
\[
\epsilon_{\uparrow}^{(0)} = - U/2,\qquad
\epsilon_{\downarrow}^{(0)} =  U/2,\qquad
V_{\uparrow}^{(0)} = {-t^*\over\sqrt{2}} = V_{\downarrow}^{(0)}.
\]
The Anderson model for Hubbard star S(1) corresponds therefore 
to a two-site cluster \cite{note1} and the number of states
contributing to the ground-state (which has one $\uparrow$-
and one $\downarrow$-electron) is
${2\choose1}{2\choose1}=4$. The numbers of one-particle
and one-hole excited states for S(1) are all
\[
M_{(\pm1,0)}^{(1)}\ =\,2\,=\,M_{(0,\pm1)}^{(1)} 
\]
and S(2), the star of the stars, corresponds to a
5-site cluster. The ground-state is realized for
three $\uparrow$- and two $\downarrow$-electrons
(the state with
two $\uparrow$- and three $\downarrow$-electrons
is higher in energy). One has therefore
\[
M_{(-1,0)}^{(2)} = 50 = M_{(0,1)}^{(2)},\qquad
M_{(1,0)}^{(2)} = 100 = M_{(0,-1)}^{(2)}.
\]
S(3) corresponds therefore to a 151-site cluster.

Above considerations are valid for constructing
the exact ground-state and one-particle
Green's function in the antiferromagnetic state.
Effective Anderson models for S(1) and S(2)
can although be constructed for the
paramagnetic state, though with an increased
size \cite{Gros94}.

%%%%%%%%%%%%%%%%%%%%%%%%%%%%%%%%%%%%%%%%%%%%%%%%%%%%%%%%%%%
\begin{table}
\caption{The ground-state expectation values for the
number of doubly occupied sites, $\langle d\rangle$,
the local moment, $\langle m\rangle$ and of the kinetic
energy, $\langle T\rangle$ (in units of $t^*$),
see Eq.\ (\protect\ref{defs}). The 
values are for $U=8$ and various Hubbard stars.
$Z$ is the weight of the states retained, 
see Eq.\ (\protect\ref{sum-rule}). $L$ is the effective
cluster size used.}
\label{table1}
\begin{center}
\begin{tabular}{llllll}
\hline\noalign{\smallskip}
& $\langle d\rangle$ & $\langle m\rangle$ &$\langle T\rangle$ 
& $Z$ & $L$\\
\noalign{\smallskip}\hline\noalign{\smallskip}
S(1) & .007687 &           -.492080 & -.249824 & 1.0    & 2\\
S(2) & .007797 & \phantom{-}.491952 & -.249810 & 1.0    & 5\\
S(3) & .007807 &           -.491941 & -.249896 & 0.99993& 7\\
S(4) & .007807 &           -.491940 & -.249901 & 0.99995&15\\
\noalign{\smallskip}\hline
\end{tabular}
\end{center}
%\vspace*{5cm}  % with the correct table height
\end{table}
%%%%%%%%%%%%%%%%%%%%%%%%%%%%%%%%%%%%%%%%%%%%%%%%%%%%%%%%%%%

%%%%%%%%%%%%%%%%%%%%%%%%%%%%%%%%%%%%%%%%%%%%%%%%%%%%%%%%%%%%
\section{Discussion}
\label{discussion}

In view of the fact that the Hilbert-space increases
exponentially with every iteration one needs to discuss
the feasibility of truncation schemes. From the 
anticommutation rule
\[
c_{i,\sigma}^{\phantom{\dagger}}
c_{i,\sigma}^{\dagger} +
c_{i,\sigma}^{\dagger}
c_{i,\sigma}^{\phantom{\dagger}} \, =\, 1
\]
for every $i=1,\dots Z$ and $\sigma=\uparrow,\downarrow$ one can
easily derive the sum-rule
\begin{equation}
\sum_{k_{\bs}}\left| a(k_{\bs},0_{(0,0)})\right|^2 +
\sum_{k_{-\bs}}\left| a(k_{-\bs},0_{(0,0)})\right|^2\, =\, 1,
\label{sum-rule}
\end{equation}
which hold for both $\bs=(1,0)$ and $\bs=(0,1)$.
Truncation schemes become feasible, when an
exponentially large number of the
matrix-elements (\ref{V_sigma}) become small
due to the sum-rule (\ref{sum-rule}). An indication
of whether this is the case or not may be seen
by studying the distribution of the
$|a(k_{\pm\bs},0_{(0,0)})|^2 $ for S(2), which 
is given in Fig.\ \ref{Fig2}.

Due to the formation of a local moment the
matrix-elements shown in Fig.\ \ref{Fig2}
are different for $\sigma=\,\downarrow$
(solid bars) and 
$\sigma=\,\uparrow$ (dashed bars). Note,
that Fig.\ \ref{Fig2} can also be interpreted
as the one-particle Green's function of S(2) \cite{Gros94}.
We observe that only a limited number of the transition
matrix elements shown in Fig.\ \ref{Fig2} has an appreciable
weight. The number of relevant poles increases 
with cluster size.

%%%%%%%%%%%%%%%%%%%%%%%%%%%%%%%%%%%%%%%%%%%%%%%%%%%%%%%%%%%
\begin{table}
\caption{The same as in table \protect\ref{table1},
for $U=4$.}
\label{table2}
\begin{center}
\begin{tabular}{llllll}
\hline\noalign{\smallskip}
& $\langle d\rangle$ & $\langle m\rangle$ &$\langle T\rangle$ 
& $Z$ & $L$\\
\noalign{\smallskip}\hline\noalign{\smallskip}
S(1) & .029127 &           -.467646 & -.495066 & 1.0    &2\\
S(2) & .030323 & \phantom{-}.465713 & -.493618 & 1.0    &5\\
S(3) & .030728 &           -.465060 & -.495726 & 0.99994&15\\
\noalign{\smallskip}\hline
\end{tabular}
\end{center}
%\vspace*{5cm}  % with the correct table height
\end{table}
%%%%%%%%%%%%%%%%%%%%%%%%%%%%%%%%%%%%%%%%%%%%%%%%%%%%%%%%%%%

We now consider in detail the ground-state
properties of various Hubbard stars for $U=8,\ 4$ and $U=2$
(see Table \ref{table1}, \ref{table2} and
\ref{table3}). For S(1) and S(2) we have constructed
the exact ground-state wavefunctions and calculated
the ground-state expectation values of the 
doubly-occupancy $\langle d\rangle$,
the local moment, $\langle m\rangle$ and of the kinetic
energy, $\langle T\rangle$, with (see Eq.\ (\ref{S_n})
\begin{eqnarray}
d=f_\uparrow^\dagger f_\uparrow^{}
   f_\downarrow^\dagger f_\downarrow^{},\qquad
m=(f_\uparrow^\dagger f_\uparrow^{}
  -f_\downarrow^\dagger f_\downarrow^{})/2
\nonumber\\
T={t^*\over\sqrt{2}}\sum_{\sigma=\uparrow,\downarrow}
\left[
f_{\sigma}^\dagger A_{\sigma}^{\phantom{\dagger}}
+A_{\sigma}^\dagger f_{\sigma}^{\phantom{\dagger}}
\right].
\label{defs}
\end{eqnarray}
Since S(3) corresponds to a cluster with $L=151$ sites
we cannot diagonalize S(3) exactly. For $U=2$ and $U=4$
we have approximated S(3) with a $L=14+1$ site cluster,
i.e.\ we have retained only the largest 14 transition
matrix elements (per spin)
$\, a(k_{\bs},0_{(0,0)})\, $ in Eq.\ (\ref{eq:adagger})
(all other matrix elements are set to zero). 
We also give
in  Table \ref{table2} and \ref{table3} the accuracy of
this approximation, i.e.\ the contribution $Z$ of the
largest 14 matrix elements per spin to the sum-rule
Eq.\ (\ref{sum-rule}). We observe that the truncation
is better for larger values of $U$. For $U=8$ it is
possible to approximate S(3) already with by a
$L=6+1$ site cluster and S(4) by a $L=14+1$ site
cluster, see Table \ref{table1}.

By setting the transition matrix elements to certain
excited states on the periphery to zero a
{\it variational} approximation is obtained. It turns
out that all the neglected excitations (those with
very small $\, a(k_{\bs},0_{(0,0)}))$  are high in
energy in the sense that they are either above the
upper Hubbard band or below the lower Hubbard band,
they contribute only to the tails of the respective Hubbard bands.
One can therefore estimate the contribution of these
excitations to the ground-state of S(3) by perturbation
theory, their weight is $<10^{-3}$ for $U=2$ and $<10^{-4}$ 
for $U=4,8$. Inspecting the data presented in the Table
\ref{table1}, \ref{table2} and \ref{table3} 
for various S(n) one sees that convergence with $n$
is good, especially for $U=4$ and $U=8$. For $U=8$
the data has converged to within $0.01\%$ for
$\langle m\rangle$, to within $0.1\%$ for $\langle T\rangle$ 
and to within $1\%$ for $\langle d\rangle$.

In summary we have shown how the finite dimensional, minimal
interacting Hilbert-Space of the Hubbard model on the infinite-D
Bethe-lattice can be constructed iteratively on the basis of symmetry
adapted transition operators. We explicitly provided the close
relationship to the recursive construction of the one-particle Green's
function based on the mapping to a self-consistent Anderson model.  We
have discussed the feasibility of truncation schemes in the iteration
process, which are necessary due to the exponential increase of the
Hilbert-space at every step of the iteration. We have found that the
sum-rule for the transition-matrix elements leads to a natural
truncation criterion. We have applied the truncation scheme
to various Hubbard stars and estimated the ground-state 
properties of the half-filled Hubbard
model up to $0.1\%$ accuracy.

%%%%%%%%%%%%%%%%%%%%%%%%%%%%%%%%%%%%%%%%%%%%%%%%%%%%%%%%%%%
\begin{table}
\caption{The same as in table \protect\ref{table1},
for $U=2$.}
\label{table3}
\begin{center}
\begin{tabular}{llllll}
\hline\noalign{\smallskip}
& $\langle d\rangle$ & $\langle m\rangle$ &$\langle T\rangle$ 
& $Z$ & $L$\\
\noalign{\smallskip}\hline\noalign{\smallskip}
S(1) & .091221 &           -.379144 & -.905348 & 1.0    &2\\
S(2) & .092159 & \phantom{-}.372893 & -.855795 & 1.0    &5\\
S(3) & .097549 &           -.359257 & -.878246 & 0.99953&15\\
\noalign{\smallskip}\hline
\end{tabular}
\end{center}
%\vspace*{5cm}  % with the correct table height
\end{table}
%%%%%%%%%%%%%%%%%%%%%%%%%%%%%%%%%%%%%%%%%%%%%%%%%%%%%%%%%%%

%%%%%%%%%%%%%%%%%%%%%%%%%%%%%%%%%%%%%%%%%%%%%%%%%%%%%%%%%%%%


\begin{thebibliography}{}
 
\bibitem{Metzner89} W. Metzner and D. Vollhardt,
                    Phys. Rev. Lett., \textbf{62} (1989) 324.

\bibitem{MH89} E. M\"uller-Hartmann,
               Z. Phys. B, \textbf{74} (1989) 507.

\bibitem{Georges92} A. Georges and G. Kotliar,
                    Phys. Rev. B, \textbf{45} (1992) 6479.

\bibitem{Gros94} C. Gros, W. Wenzel, R. Valent\'\i, 
                 G. H\"ulsenbeck and J. Stolze,
                 Europhys. Lett., \textbf{27} (1994) 299.

\bibitem{Georges96} A. Georges, G. Kotliar, W. Krauth and M. Rozenberg,
                    Rev. Mod. Phys., \textbf{68} (1996) 13.

\bibitem{Bulla98} R. Bulla, A.C. Hewson and Th. Pruschke,
                  cond-mat/9804224.

\bibitem{Dongen91} P.G.J. van Dongen, J.A. Verg\'es and D. Vollhardt,
                   Z. Phys. B, \textbf{84} (1991) 383.

\bibitem{note1} In Ref.\ \protect\cite{Dongen91} a mapping of the
                Hubbard star to a three-site model has been given.
                For the ground-state this Hilbert-space can be
                reduced to a two-site model.

%\bibitem{Fledder96} A. Fledderjohann, M. Karbach and K.-H. M\"utter,
%                    Phys. Rev. B, \textbf{53} (1996) 11 543.

\bibitem{Gros97} C. Gros, W. Wenzel, A. Fledderjohann, P. Lemmens, 
                 M. Fischer, G. G\"untherodt, M. Weiden, 
                 C. Geibel, F. Steglich, 
                 Phys. Rev. B \textbf{55} (1997) 15048.

\end{thebibliography}
\end{document}